# Analysis of $L2_1$-ordering in full-Heusler Co$_2$FeSi alloy thin films formed by rapid thermal annealing

Yota Takamura[1,a)], Ryosho Nakane[2], and Satoshi Sugahara [1,3,4,b)]

[1]*Department of Electronics and Applied Physics, Tokyo Institute of Technology, Yokohama 226-8503, Japan*

[2]*Department of Electrical Engineering and Information Systems, The University of Tokyo, Tokyo 113-8656, Japan*

[3]*Imaging Science and Engineering Laboratory, Tokyo Institute of Technology, Yokohama 226-8502, Japan*

[4]*CREST, Japan Science and Technology Agency, CREST, Kawaguchi 332-0012, Japan*

a) Electronic mail: yota@isl.titech.ac.jp

b) Electronic mail: sugahara@isl.titech.ac.jp

ABSTRACT

The authors developed a new analysis approach for evaluation of atomic ordering in full-Heusler alloys, which is extension of the commonly used Webster model. Our model can give accurate physical formalism for the degree of atomic ordering in the $L2_1$ structure, including correction with respect to the fully disordered $A2$ structure, i.e., the model can directly evaluate the degree of $L2_1$-ordering under a lower ordering structure than the complete $B2$-ordering structure. The proposed model was applied to full-Heusler $Co_2FeSi$ alloy thin films formed by rapid thermal annealing. The film formed at $T_A$ = 800 °C showed a relatively high degree of $L2_1$-ordering of 83 % under a high degree of $B2$-ordering of 97 %.

BODY

Recently, full-Heusler alloys, such as $Co_2FeSi$[1] and $Co_2MnSi$[2,3], attract considerable attention, since they have an unique half-metallic band structure[4] with high Currie temperature. In particular, Si-containing full-Heusler alloys are a promising material for spin injector/detector of Si-based spin devices such as spin metal-oxide-semiconductor field-effect transistors (spin MOSFETs)[5-7], since they can be formed by silicidation reaction activated by rapid thermal annealing (RTA) that is commonly used formation process for metal source/drain technology in advanced CMOS devices. Recently, we reported that full-Heusler $Co_2FeSi$ alloy thin films were successfully formed by RTA using silicon-on-insulator (SOI) substrates[8].

A fully ordered atomic arrangement in full-Heulser alloys is the $L2_1$ structure, however, the partially disordered $B2$ and fully disordered $A2$ structures also exist. Half-metallicity in full-Heusler alloys is sensitive to their atomic arrangement, and degrades with deterioration of atomic ordering.[9] Therefore, characterization of atomic ordering in the $L2_1$ structure is an important concern for half-metallic full-Heusler alloys[10]. Nevertheless, the standard approach to evaluate the degree of atomic ordering in full-Heusler alloys still leaves some issues. In this paper, we developed a novel analysis approach to evaluate atomic ordering in full-Heusler alloys, extending the commonly used Webster scheme. The proposed model was applied to full-Heusler $Co_2FeSi$ alloy thin films formed by RTA.

The $L2_1$ structure of full-Heusler $X_2YZ$ alloys (such as $Co_2FeSi$) is composed of eight stacked body-centered cubic (bcc) lattices shown in Fig. 1(a). The outside sublattice that consists of the eight cubic lattices is occupied by X atoms, and the inside cubic sublattice that consists of the body-centered sites of each bcc lattice is regularly occupied by Y and Z atoms, as shown in Fig. 1(b). In full-Heusler alloys, two kinds of disordering exist. When the YZ sublattice is randomly occupied by Y and Z atoms, i.e., disordering in the YZ sublattice occurs, the ordering structure is reduced to the $B2$ type. (Hereafter, this disorder process is denoted by Y-Z disordering.) Furthermore, when disordering between the X and YZ sublattices occurs, the ordering structure is lowered to the $A$2 type. (Hereafter, this disorder process is referred to as X-YZ disordering.) In general, these ordered/disordered structures can be identified by x-ray diffraction (XRD) analysis using the following features on relations between atomic orderings and superlattice diffraction lines: Y-Z disordering extinguishes odd superlattice diffraction lines (that are defined by the index relation of $h$, $k$, and $l$ = odd numbers, e.g., (111)). Furthermore, even superlattice diffraction lines ($(h+k+l)/2 = 2n+1$, e.g., (200)) vanish under X-YZ disordering. On the other hand, fundamental diffraction lines ($(h+k+l)/2 = 2n$, e.g., (220)) are independent of the ordering structures.

The traditional analytical approach[11,12] proposed by Webster has been extensively used to evaluate the degree of atomic ordering in full-Heusler alloys. In the Webster framework, ordering features in full-Heusler alloys are divided into two factors, i.e., the degree of $B$2-ordering, and the degree of $L2_1$-ordering. The degree of $B$2-ordering, $S_{B2}^W$, represents that of X-YZ disordering (the definition will be discussed later). Although the degree of $L2_1$-ordering, $S_{L2_1}^W$, can be defined by the same manner as that of $B$2-ordering (also discussed later), the disordering parameter $\alpha^W$ that is defined by a fraction of Y atoms on the Z sites is generally used to evaluate $L2_1$-ordering. Both $S_{L2_1}^W$ and $\alpha^W$ represent an index of the degree of Y-Z disordering, however, the definition and physical meaning are different. $S_{B2}^W$, $S_{L2_1}^W$, and $\alpha^W$ can be easily determined from XRD intensity ratios between superlattice and fundamental lines.[11,12] (The corrected formula to deduce $\alpha^W$ is presented in Ref. 12.) The Webster model can give accurate formalism for the degree of $B$2-ordering [11]. On the other hand, $\alpha^W$ is given under the hypothesis of the complete $B$2 structure of $S_{B2}^W$ = 100 %. Namely, $\alpha^W$ in the Webster model has no influence of X-YZ disordering. In other words, when $S_{B2}^W$ is less than 100 %, the physical meaning of $\alpha^W$ is ambiguous. It should be noted that although $\alpha^W$ represents the fraction of Y atoms on the Z sites, the number of Y atoms on the Z sites also depends on the degree of $B2$-ordering.

Extending the Webster model, we developed physical formalism for the degree of $L2_1$-ordering including correction from the degree of $B2$-ordering. Table 1 shows a definition of the number of atoms on each site (in unit of a $Co_2Fe_1Si_1$ molecule). In the table, $2x$ ($0.5 \leq x \leq 1$) and $y$ ($x/2 \leq y \leq x$) are the number $n_{Co\,on\,X}$ of Co atoms on the X sites, and the number $n_{Fe\,on\,Y}$ of Fe atoms on the Y sites, respectively. The number of Si atoms on the Z sites is also $y$, where we assumed that X-Y and X-Z disorderings occur equiprobably, which is the same as the assumption of the Webster model[11,12]. The other values (e.g., the number of Si atoms on the Y sites) can be easily deduced under the assumption of the stoichiometric composition. The proposed model can express two X-YZ and Y-Z disorderings simultaneously. The degree of $B2$-ordering, $S_{B2}$, is defined as follows:

$$S_{B2} = \frac{n_{Co\,on\,X} - n_{Co\,on\,X}^{random}}{n_{Co\,on\,X}^{full\text{-}order} - n_{Co\,on\,X}^{random}} = 2x - 1, \qquad (1)$$

where $n_{Co\,on\,X}^{random}$ (= 1) is the number of Co atoms on the X sites for the most random distribution of Co atoms in the $A2$ structure, and $n_{Co\,on\,X}^{full\text{-}order}$ (= 2) is the number of Co atoms on the X sites in the fully-ordered X sublattice of the perfect $B2$ structure. In practice, $S_{B2}$ can be obtained from XRD measurements using the following relation between even superlattice and fundamental lines,

$$\frac{I_{200}}{I_{220}} = (2x-1)^2 \frac{I_{200}^{full\text{-}order}}{I_{220}^{full\text{-}order}} = S_{B2}^2 \frac{I_{200}^{full\text{-}order}}{I_{220}^{full\text{-}order}}, \qquad (2)$$

where $I_{200}/I_{220}$ is an experimentally obtained intensity ratio between (200) and (220) diffraction lines, and $I_{200}^{full\text{-}order}/I_{220}^{full\text{-}order}$ is the (calculated) ideal diffraction intensity ratio for the perfect $B2$ structure (that is whole the X sites are occupied only by Co atoms). The expression of $S_{B2}$ in our model corresponds to that in the Webster model ($S_{B2} = S_{B2}^W$). The degree of $L2_1$-ordering, $S_{L2_1}$, can be also defined by the following relation,

$$S_{L2_1} = \frac{n_{\text{Fe on Y}} - n_{\text{Fe on Y}}^{\text{random}}}{n_{\text{Fe on Y}}^{\text{full-order}} - n_{\text{Fe on Y}}^{\text{random}}} = \frac{2y - x}{2 - x}, \qquad (3)$$

where $n_{\text{Fe on Y}}^{\text{random}}$ ($= x/2$) is the number of Fe atoms on the Y sites for the most random distribution of Fe atoms in the YZ sublattice, and $n_{\text{Fe on Y}}^{\text{full-order}}$ is that for the fully ordered YZ sublattice. The codomain of $S_{L2_1}$ depends on $S_{B2}$ and it is limited to a range of $0 \leq S_{L2_1} \leq \frac{1 + S_{B2}}{3 - S_{B2}}$. Note that $S_{L2_1}$ can take a finite value even under $S_{B2} = 0$, since $L2_1$-ordering portion without Y-Z disordering can remain when the X sublattice are fully disordered (a half amount of whole the Fe and Si atoms is required for the fully disordered X sublattice). $S_{L2_1}$ can be obtained from the following relation,

$$\frac{I_{111}}{I_{220}} = (2y - x)^2 \frac{I_{111}^{\text{full-order}}}{I_{220}^{\text{full-order}}} = \left[ S_{L2_1} \left( \frac{3 - S_{B2}}{2} \right) \right]^2 \frac{I_{111}^{\text{full-order}}}{I_{220}^{\text{full-order}}}. \qquad (4)$$

This relation can be easily deduced by the ordinary procedure[11] from crystal structure factors ($F_{111} = 4|f_Y - f_Z|$ and $F_{220} = 4|2f_X + f_Y + f_Z|$) and average atomic scattering factors at each site ( $f_X = 2xf_{\text{Co}} + (1-x)f_{\text{Fe}} + (1-x)f_{\text{Si}}$, $f_Y = (1-x)f_{\text{Co}} + yf_{\text{Fe}} + (x-y)f_{\text{Si}}$, and $f_Z = (1-x)f_{\text{Co}} + (x-y)f_{\text{Fe}} + yf_{\text{Si}}$). It should be noted that $S_{L2_1}$ cannot be determined only by the odd superlattice line, and that $S_{L2_1}$ also depends on the degree of $B2$-ordering. In the Webster framework, $S_{L2_1}^{\text{W}}$ is given by $S_{L2_1}^{\text{W}} = 1 - 2\alpha^{\text{W}}$ using the general definition shown in the second part of Eq. (3) and it is given by the relation of $I_{111}/I_{220} = S_{L2_1}^{\text{W}\,2} I_{111}^{\text{full-order}} / I_{220}^{\text{full-order}}$. When $S_{B2} = 100\%$ ($x = 1$), our $S_{L2_1}$ formula ($= 2y - 1$) is identical with Webster's $S_{L2_1}^{\text{W}} (= 1 - 2\alpha^{\text{W}})$ (since $x = 1$, Webster's disordering parameter is expressed by $\alpha^{\text{W}} = 1 - y$). Moreover, $S_{L2_1}$ can evaluate the degree of $L2_1$-ordering under a particular $S_{B2}$ value of $S_{B2} <$

100 %.

Figure 2 shows $S_{L2_1}$ as a function of $S_{B2}$ with various diffraction intensity ratios $I_{111}/I_{111}^{\text{full-order}}$. The solid curves show $S_{L2_1}$ in our model, and the solid circles show $S_{L2_1}^{\text{W}}$ in the Webster framework that can be only plotted on $S_{B2}$ = 100 %. $S_{L2_1}$ and $S_{L2_1}^{\text{W}}$ increase with increasing $I_{111}/I_{111}^{\text{full-order}}$. Although $S_{L2_1}$ are identical with $S_{L2_1}^{\text{W}}$ when $S_{B2}$ = 100 %, $S_{L2_1}$ decreases with decreasing $S_{B2}$. It should be emphasized that $S_{L2_1}$ can be evaluated for overall $S_{B2}$, whereas $S_{L2_1}^{\text{W}}$ is only defined at $S_{B2}$ = 100 % ($x$ = 1).

The developed model was applied to RTA-formed $Co_2FeSi$ thin films[8]. Recently, we reported that full-Heusler $Co_2FeSi$ alloy thin films were successfully formed by RTA-induced silicidation reaction utilizing SOI substrates. Detailed experimental procedure was described in Ref. 8. After the chemical cleaning of a SOI substrate, Co and Fe films were deposited on the SOI layer surface in an ultrahigh vacuum. Subsequently, the silicidation was performed by RTA in $N_2$ atmosphere. RTA temperature $T_A$ was varied in a range between 600-800 °C. The thicknesses of Co and Fe layers were determined so that the stoichiometric composition was achieved. The depth profiles of constituent elements were observed by secondary ion mass spectroscopy (SIMS) with $MCs^+$ technique[13], whose signals were calibrated by Rutherford backscattering (RBS) and particle induced x-ray emission (PIXE) measurements. The composition of the sample formed at $T_A$ = 800 °C was homogeneous and stoichiometric. The concentrations of Co, Fe, and Si in the $Co_2FeSi$ film were 49 %, 26 %, and 25 %, respectively.

The crystallographic features of the samples were characterized by XRD. XRD patterns for the samples formed above $T_A$ = 700 °C showed $Co_2FeSi$(220) and $Co_2FeSi$(440) diffraction in a range of 20 ° ≤ 2$\theta$ ≤ 120 °. Pole figure analysis for the (220) fundamental diffraction revealed that the RTA-formed $Co_2FeSi$ thin films were highly (110) oriented columnar polycrystalline. This result was consistent with our transmission electron microscopy (TEM) observation.

In-plane XRD measurements were performed to evaluate intensity ratios between superlattice and fundamental diffraction lines that characterize the degree of $B2$- and $L2_1$-orderings. Since the RTA-formed $Co_2FeSi$ films were (110)-oriented columnar polycrystalline, the three important diffraction lines of (-111), (002), and (2-20) can be detected simultaneously in in-plane measurements. As a result, accurate intensity relations were obtained without complicated corrections. Figure 3 shows XRD patterns measured with in-plane configuration for the samples formed at $T_A$ ranging from 600 to 800 °C. Above $T_A$ = 650 °C, the three (-111), (002), and (2-20) diffraction peaks were clearly observed, indicating the formation of the $L2_1$-ordered structure. The intensity ratios $I_{002}/I_{2-20}$ between (002) and (2-20) lines increased with increasing $T_A$ up to $T_A$ = 750 °C, and they were saturated above $T_A$ = 750 °C. On the other hand, the intensity ratios $I_{-111}/I_{2-20}$ between (-111) and (2-20) lines increased with increasing $T_A$ even for $T_A$ > 750 °C. The degree of atomic orderings in the films was estimated using these intensity ratios. Figure 4 shows $S_{B2}$, $S_{L2_1}$, and $S_{L2_1}^W$, as a function of $T_A$. In order to deduce these indices, the intensity ratios $I_{002}^{\text{full-order}}/I_{2-20}^{\text{full-order}}$ and $I_{-111}^{\text{full-order}}/I_{2-20}^{\text{full-order}}$ shown in Eqs. (2) and (4), respectively, were evauated from International tables for crystallography volume C and D with the correction of multiplicity. The sample formed at $T_A$ = 800°C showed a relatively high degree of atomic ordering of $S_{L2_1}$ = 83 % under $S_{B2}$ = 97 %. It is clearly shown that $S_{L2_1}^W$ was overestimated in comparison with $S_{L2_1}$, and that the deviation from $S_{L2_1}$ increased with decreasing $S_{B2}$ as described in Fig. 4.

In summary, we developed a new analysis approach for characterization of atomic orderings in full-Heusler alloys, extending the Webster model. The proposed model can express accurate physical formalism for $B2$- and $L2_1$-orderings.  The model was applied to full-Heusler $Co_2FeSi$ alloy thin films formed by RTA.  The film formed at $T_A$ = 800°C showed a relatively high degree of $L2_1$-ordering of 83 % under a high degree of $B2$-ordering of 97 %.

The authors would like to thank Prof. H. Munekata, Tokyo Institute of Technology, Prof. M. Tanaka and S. Takagi, The University of Tokyo.  In-plane XRD measurements were performed by Rigaku Corporation.

Figure captions

Fig. 1 Schematic view of (a)the $L2_1$ structure and (b)the inside cubic sublattice.

Table 1 Definition of the number of atoms at each site in full-Heulser $Co_2FeSi$ alloy (in unit of a $Co_2Fe_1Si_1$ molecule).

Fig. 2 Degree of $L2_1$-ordering, $S_{L2_1}$, as a function of the degree of $B2$-ordering, $S_{B2}$, with various diffraction intensity ratios $I_{111}/I_{111}^{\text{full-order}}$. The solid curves show $S_{L2_1}$, and the solid circles $S_{L2_1}^{W}$ in the Webster framework. The filled region represents the codomain of $S_{L2_1}$ given by $0 \leq S_{L2_1} \leq (1+S_{B2})/(3-S_{B2})$.

Fig. 3 XRD patterns measured with in-plane configuration for the $Co_2FeSi$ films formed at $T_A$ ranging from 600 to 800 °C

Fig. 4 Degree of $B2$- and $L2_1$-orderings, $S_{B2}$, and $S_{L2_1}$, as a function of RTA temperature $T_A$. The thin solid curve reprents $S_{B2}$, the thick solid curve $S_{L2_1}$. The degree of $L2_1$-ordering in the Webster framework, $S_{L2_1}^{W}$, is also shown by the dashed curve in the figure.

|  | X sites | Y sites | Z sites |
|---|---|---|---|
| No. of Co atoms | $2x$ | $1-x$ | $1-x$ |
| No. of Fe atoms | $1-x$ | $y$ | $x-y$ |
| No. of Si atoms | $1-x$ | $x-y$ | $y$ |

Table 1 Takamura *et al*.

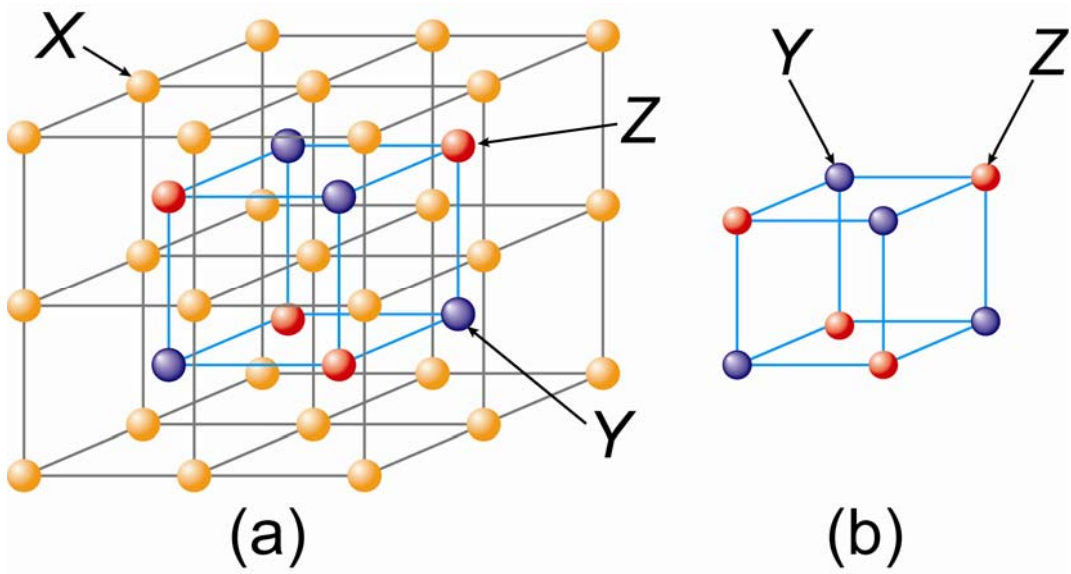

Figure 1 Takamura *et al*.

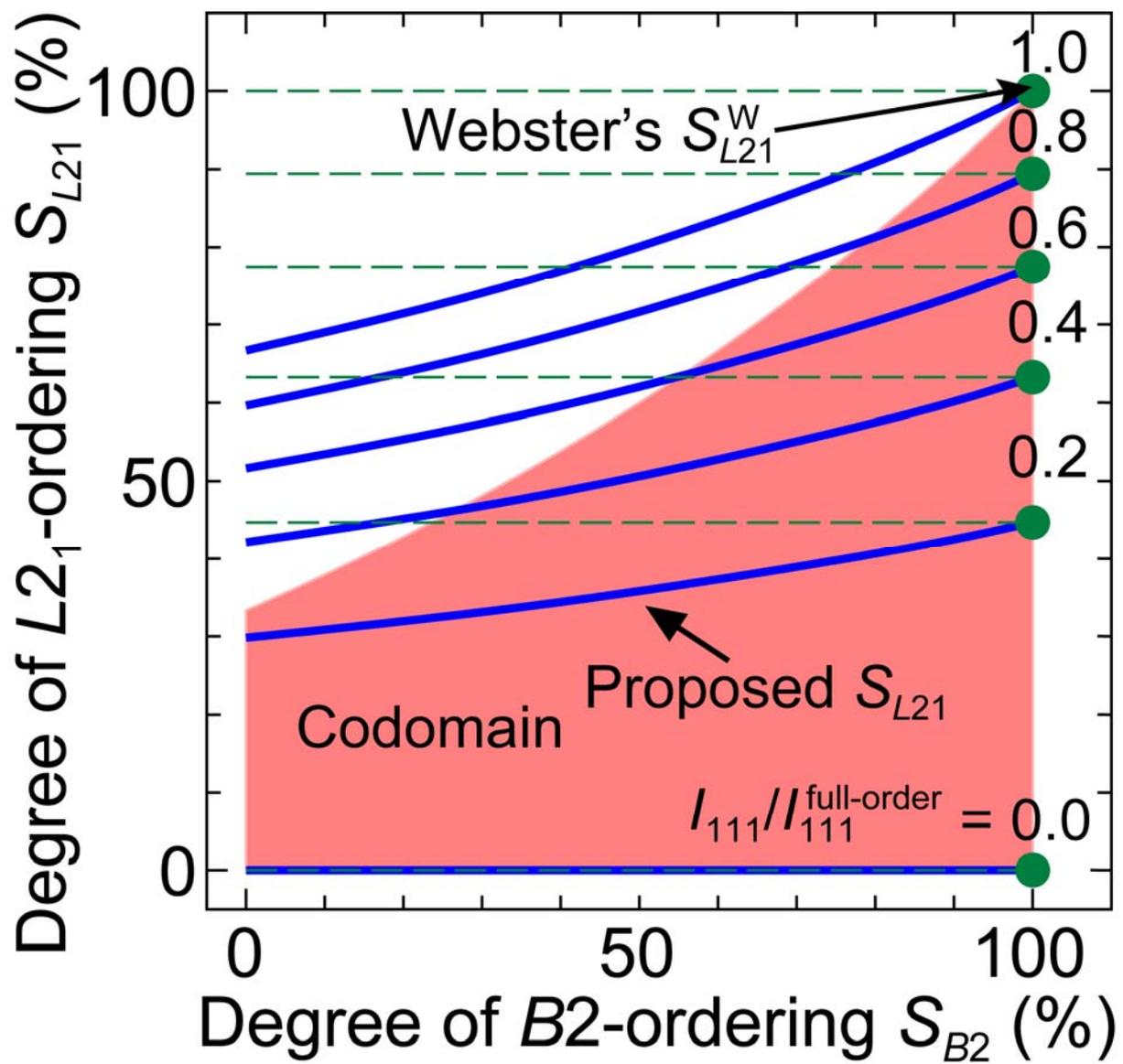

Figure 2 Takamura *et al*.

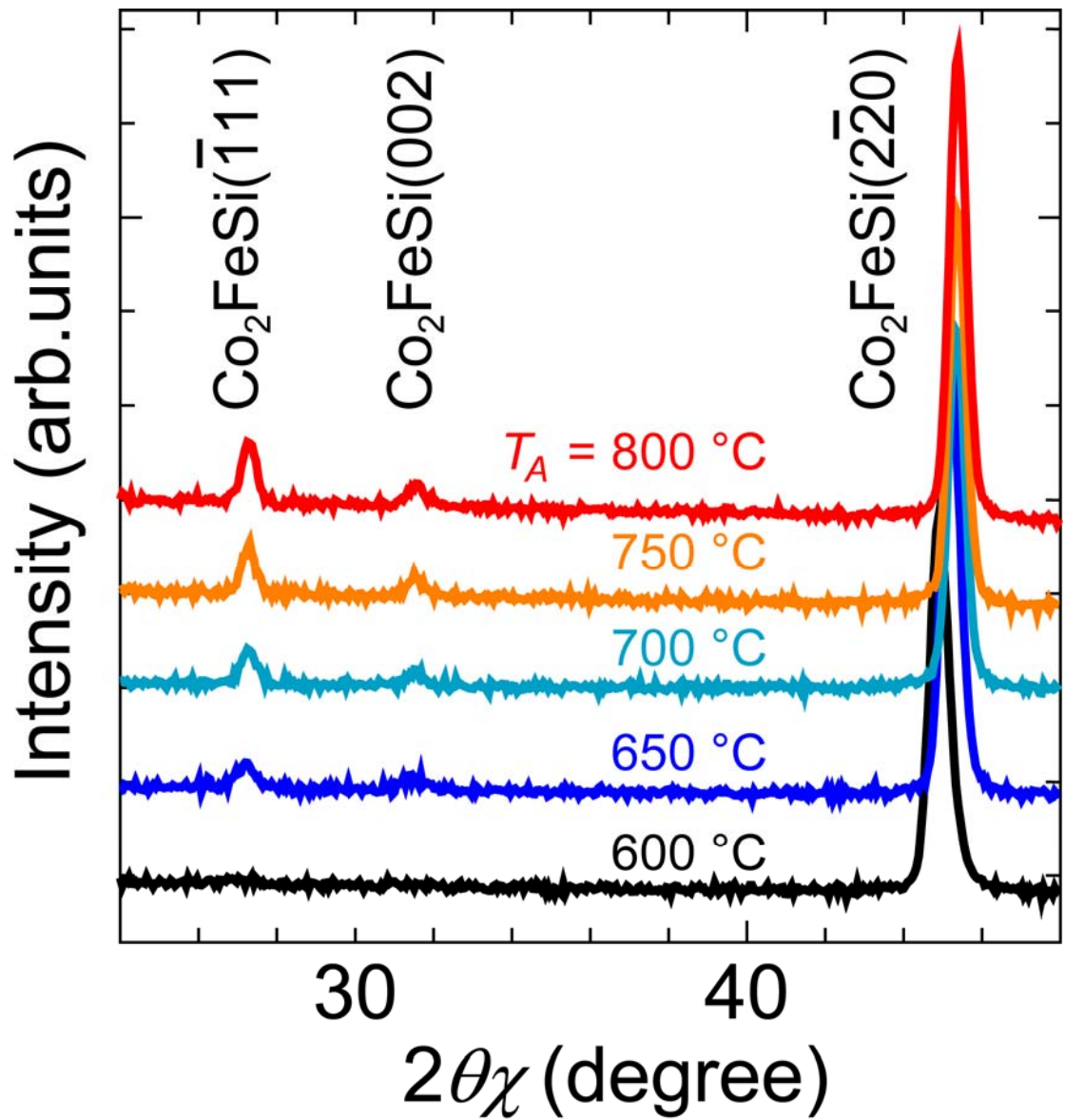

Figure 3 Takamura *et al*.

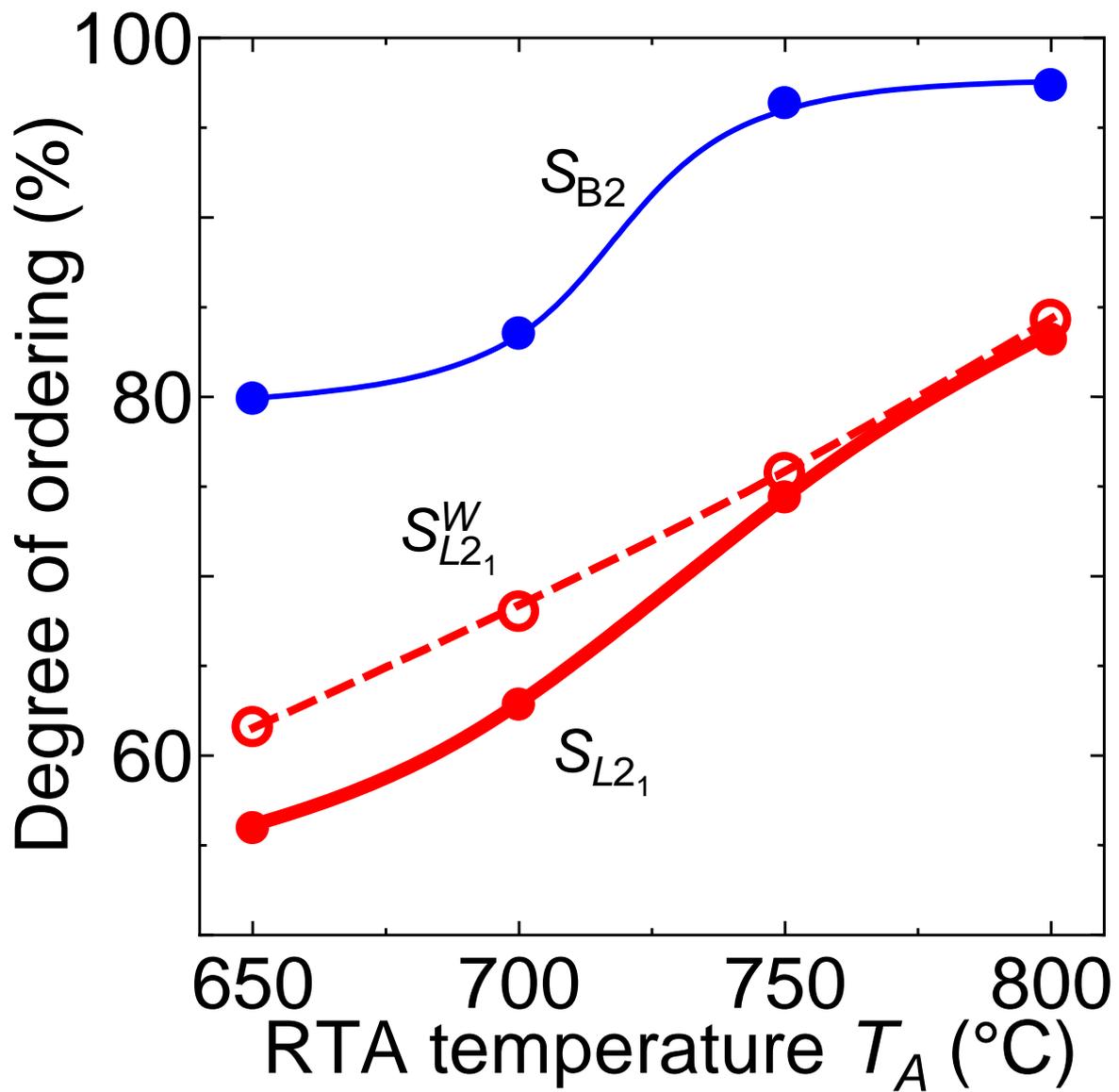

Figure 4 Takamura *et al*.